\documentclass[times,epsf,overcite]{AJParticle}
\usepackage{graphicx}
\usepackage{cite}
\begin{document}
\begin{center}
{\large\bf Cavity Solitons}\\*[0.25cm]
W.J. FIRTH and G.K. HARKNESS\\*[0.0cm]
{\small\it Department of Physics and Applied Physics,
University of Strathclyde,\\
Glasgow G4 0NG, Scotland}\\*[0.5cm]
\parbox{10cm}{\small A driven optical cavity containing a nonlinear
medium can support stable, soliton-like objects, bright non-diffracting
spots of light on a uniform background.  Importantly, the nonlinear
medium need not support ordinary (propagating) spatial solitons, and
cavity solitons have been both predicted and observed in a saturable
absorber, and predicted to even exist in self-defocusing media. The
history, present status, and physical interpretation of such cavity
solitons is reviewed. Recent research predicts cavity solitons in
semiconductor cavity models with properties interesting for
applications to optical information processing.  Cavity solitons are
natural `bits' for spatial information, possessing both robustness and
plasticity, and some potential applications of their unique properties
are mentioned.}
\end{center}

\section*{Introduction}
Optical solitons are pulses (or beams) of light in which nonlinearity
counter-balances dispersion (or diffraction), leading to a robust
structure which propagates without change of form.  Strictly speaking,
true solitons are exact solutions of {\em integrable} nonlinear partial
differential equations, but in nonlinear optics mathematical nicety is
often foregone in favour of physical robustness.  Thus the term
`soliton' is increasingly used for any pulse or beam in which
dispersion and/or diffraction is compensated {\em on the average} by
nonlinearity.

Many interesting spontaneous spatial structures have recently been
predicted and/or observed in nonlinear optical systems\cite {CSF}.
Among such structures are localised bright spots in driven optical
cavities
\cite{Mol,Roz,McD,Wab,Mc2,Tli,OBH,KC,PhSc,EPL,CSFtli,CSFscr,Longhi,Mit,PASSrep,MilanPRL,JenaPRL,JenaPRE,Stal,TMH,JenaVic,TlidiPinos}.
They share some properties with spatial solitons, counterparts of the
soliton pulses being developed as `bits' for long-haul fibre-optic
communications, and we will  refer to them as {\em cavity solitons}.
Such structures could be natural `bits' for parallel processing of
optical information, especially if they exist in semiconductor
micro-resonators.  Recent work by the authors and others
\cite{PASSrep,MilanPRL,JenaPRE} are very encouraging in that respect.

\begin{figure}[htbp]
\centering
    \includegraphics[width=0.75\textwidth]{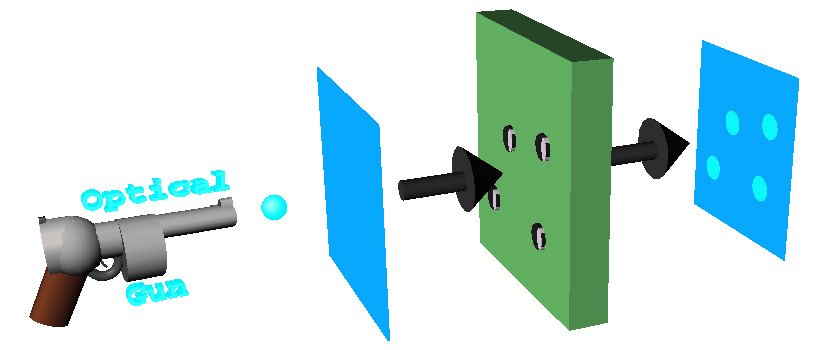}
\caption{ \label{fig:gun}
Schematic representation of an optical memory system
based on localised states in a driven optical cavity.}
\end{figure}

In this paper we briefly review the history and underlying nonlinear
optics of cavity solitons.  We then describe in more detail some models
which show these soliton-like structures. It is somewhat paradoxical
that stable cavity solitons can exist in media with properties
different from, even opposite to, those required for Kerr solitons. We
therefore consider several candidate physical interpretations of these
structures:  `soliton-in-a-box', `self-trapped switching wave',
`pattern element', and `local nonlinear resonance'.  We examine the
perturbation eigenmodes of the cavity solitons in a sample system, to
see whether their shape and nature helps us to reach a global
interpretation of these solitons.  These eigenmodes also give
information on the soliton's stability and response to external
influences such as noise, neighbouring solitons or phase gradients of
the holding (or {\em control}) field. This leads a discussion of the
applications of cavity solitons, which have also been called `optical
bullet-holes' (OBH) because they can be created by localised pulses
(bullets) of light\cite{OBH}, and can thus be used to store images or
information~-- Fig.~\ref{fig:gun} shows a cartoon version of such a
system.  The ability to control and manipulate these
solitons offers potential advantages over competing systems, and we
speculate about device ideas which might capitalize on these advantages.

Rather than an exhaustive review of the field, this article is
subjective, concentrating on our own work.  We hope, however, that it
succeeds as an overview, if not a review, and we have tried to use our
own work primarily to illustrate points of general relevance, without
dwelling on their detailed description.


\section*{History}
The story of cavity solitons can perhaps be divided into three Ages:
the Stone Age, the Space Age, and the Information Age.  The Stone Age
lasted roughly until 1990.  During it a few pioneers\cite{Mol,Roz,McD}
considered relevant spatial phenomena in cavities.  A special journal
issue\cite{JOSABsi} ended that stage, and roughly marks the transition
into the Space Age.  In the Space Age several favourable circumstances
combined to stimulate a huge expansion of interest in spatio-temporal
nonlinear optics.  Workstation development made dynamical simulation of
two-dimensional field patterns widely available.  In addition, simple
mean-field models of such systems provided a convenient framework which
readily matched on to studies of pattern formation in fluids and other
fields.  A later special issue, in 1994, \cite{CSF} shows how
dramatically the field changed in just a few years.  This special issue
includes some papers predicting mean-field cavity
solitons\cite{CSFtli,CSFscr}, a trickle which has become a flood in the
last few years.  As yet the Information Age has barely dawned: in it
the emphasis will switch from mere existence of cavity solitons to
their engineering and applications.

\begin{figure}[htbp]
\centering
    \includegraphics[width=0.90\textwidth]{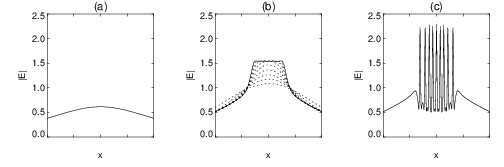}
\caption{ \label{fig:jerry}
Gaussian beam switching in an OB cavity.  The central section of a
smooth broad beam (a) switches; a switching wave then forms and moves
out (dotted curves in (b); the high-intensity central region then
breaks up into spikes (after Moloney {\em et al.}\cite{Mol}).}
\end{figure}

The history of optical cavity solitons probably began with the seminal
paper of Moloney and co-workers\cite{Mol}, who used split-step
FFT methods to simulate transverse effects in optical
bistability (OB)\cite{Gibbsbook,LALart}.  The model system was a ring
cavity, driven by a gaussian beam, and containing a self-focusing
Kerr-like medium. The field was propagated around the cavity, and added
coherently to the driving beam at the input beam-splitter. The
simulation was one-dimensional (1D), i.e. the intra-cavity field was
described by $E_n(x,z)$ with $n$ counting the cavity round-trips. When
the input field was ramped up to exceed the OB switch-up threshold, the
beam centre switched, and a {\em switching wave} moved out, switching
up most of the beam (Fig.~\ref{fig:jerry}).  Then something
unexpected happened:  a new instability.  The interface between the
`on' and `off' domains spawned what would now be termed a modulational
instability (MI) of the `on' region, which broke up into a set of
distinct peaks\cite{Mol}.  These were interpreted as a group of
spatial solitons circulating in the medium, perturbed by the output
coupling losses and sustained by the input field.  Here,
therefore, the  model was clearly `soliton-in-a-box'.  Such an
interpretation implies that only a medium which could sustain solitons
in the bulk can sustain solitons in a cavity.

The original cavity soliton work\cite{Mol}, however, considered a beam 
either with no solitons, or full of solitons: just two states, and therefore only a one-bit memory in applications terms.
McDonald and Firth\cite{McD} showed that it was possible to make the
individual solitons independently switchable by using a pump beam with
a spatially-varying amplitude.  They modeled a 20-bit memory of
this kind, and also showed that it was possible to switch solitons
`off' as well as `on' (with an out-of-phase address pulse)\cite{Mc2}.

The other key Stone Age pioneer was Rosanov, who from study of
switching waves developed the idea of `diffractive autosolitons'
\cite{Roz}, originally in the context of OB.  Switching waves between
co-existent stable states are known in many fields, such as
reaction-diffusion systems.  Purely diffusive switching waves have
monotonic profiles, and the more stable state simply wipes out the less
stable, as was shown in an OB model with pure diffusion\cite{FG}.  When
diffraction is present, however, the switching wave typically has
ripples.  These can trap other switching waves, and thus two can trap
each other.  Or, in 2D, one switching wave can bend around and close on
itself forming a stable island of one phase surrounded by the other~--
a diffractive autosoliton (DAS)\cite{Roz}.  Here, then, is a second
physical interpretation of cavity solitons: self-trapped switching
waves. Note that here there is no requirement for bulk solitons: indeed
Rosanov showed that DAS can occur even in a saturable absorber.  Nor is
OB required: one can have switching waves, e.g. between homogeneous and
patterned states, even in the absence of OB.

Rosanov has made many significant contributions to OB and related
fields, too many to describe here: the interested reader is referred to
his review article\cite{RozPO}.  He has also investigated DAS in
lasers\cite{lasDAS}, in which context we remark that we will not deal
here with cavity solitons in lasers, nor with the  self-imaging
oscillators in which {\em single} soliton-like structures have been
experimentally observed\cite{Tar,Saff}.


\section*{Mean-Field Models and Cavity Solitons}
We now enter the Space Age, characterised by so-called mean-field
cavity models, in which alternation of propagation around the cavity
with coherent addition of the input field is replaced by a single
partial differential equation with a driving term.  In the context of
spatial pattern formation, this approach is ascribed to the seminal
paper of Lugiato and Lefever\cite{LL}.  Here we give a heuristic
derivation of their LL equation, starting from the Nonlinear
Schr\"odinger Equation (NLS).  This basically means we are going to
consider `boxing-in' a spatial soliton.

Spatial solitons, in a strict sense, exist only for the case of a
single transverse dimension, which corresponds to propagation in a
planar waveguide.  This system is describable to a good approximation
by the 1D NLS, one of the best studied and understood of all nonlinear
partial differential equations:
\begin{eqnarray}
i{{\partial E} \over {\partial z}}+{1 \over 2}{{\partial ^2E} \over
{\partial x^2}}+\left|
E \right|^2E=0
\label{Sol.1}
\end{eqnarray}
Here the first and second order derivatives describe propagation and
diffraction respectively, while  the term in $|E|^2$ describes the Kerr
effect.  This dimensionless form of the NLS is able to describe
nonlinear waves in many areas of science\cite{Newell}. It has a set of
`soliton' solutions, which are filaments or pulses of hyperbolic secant
shape.  The simplest of these is the `single soliton', which in these
scaled units has the rather simple formula:
\begin{eqnarray}
E(z,x)={\mathrm sech} (x)e^{iz/2}
\label{Sol.2}
\end{eqnarray}
Note that the amplitude and shape propagate without distortion or
dispersion, because the z-dependence is just a phase rotation. Though
the NLS is idealised, in real optical media soliton-like solutions
persist when the NLS is perturbed by inevitable complications such 
as gain or absorption loss.

A further important idealisation of the NLS is that it assumes an
{\em infinite} nonlinear medium.  Real nonlinear optical media have
finite dimensions and, except in glass fibre, solitons can rarely
propagate more than a few centimetres before running out of material.
This, perhaps, makes it natural to put mirrors around the medium,
confining the soliton into a finite slab of material.  With perfect
reflection and zero absorption, one could indeed confine a soliton in a
box. Real mirrors and materials are lossy, but we can make
good the loss by `feeding' the caged soliton with an input field.  We
are thus led to consider a perturbed NLS:
\begin{eqnarray}
i{{\partial E} \over {\partial t}}+{1 \over 2}{{\partial ^2E} \over
{\partial x^2}}+\left|
E \right|^2E=i\epsilon(-E - i \theta E + E_{\mathrm in})
\label{LLeq}
\end{eqnarray}
We have added three terms on the right side of the NLS, all small if
$\epsilon$ is. The first is just a linear loss ($\epsilon>0$), and the
last is the driving field $E_{\mathrm in}$ needed to sustain $E$
against that loss.  Less obvious is the middle term, in $\theta$, but
we must remember that coherent light confined between mirrors lies
within an optical cavity, and so the response to the driving field will
strongly depend on whether or not it is in resonance with the cavity.
Hence, therefore, we need $\theta$, the {\em cavity mistuning} (if we
ignore the {\em left} side of (\ref{LLeq}), then $E=E_{\mathrm in}/(1
+i\theta)$, showing a Lorentzian resonance in $\theta$).  There is one
further change from the NLS:  propagation (in $z$) has now been
replaced by evolution (in $t$).  This is natural: the soliton is now in
a box, and not going anywhere.

In the limit $\epsilon \rightarrow 0$ Eq. (\ref{LLeq}) recovers the
NLS, with a soliton solution of sech-profile in $x$, time independent
except for a phase rotation.  We might expect, therefore, that for
finite $\epsilon$ it has sech-like {\em cavity soliton} solutions
for suitable $E_{\mathrm in}$, and indeed it has.  It is usual to
consider $E_{\mathrm in}$ to be a plane wave, independent of
$x$ and $y$, in which case the soliton sits on a homogeneous non-zero
background field $E_s$.

If we instead set $\epsilon = 1$, as we henceforth do, we get the
Lugiato-Lefever (LL) equation, which was originally introduced
\cite{LL} as a model for pattern formation.  Note that in the LL
equation time $t$ is scaled to the cavity loss time.  Having removed
$\epsilon$ in this way, to recover the NLS we should now take the limit
$\theta \rightarrow \infty$.  The LL equation can also be considered as
a `mean-field' model for OB with transverse effects. Mean-field because
it is usually derived by assuming a high finesse, so that the cavity
field is approximately constant along the cavity axis. The high finesse
allows the Airy function response of the cavity to be approximated by a
single longitudinal mode, giving the
 Lorentzian resonance mentioned above.  For a plane-wave pump, the
plane-wave cavity field obeys $E_s=E_{\mathrm in}/[1
+i(\theta-|E_s|^2)]$, which is three-valued for $\theta \ge\sqrt{3}$,
showing that the model exhibits OB\cite{LL}. In fact, Lugiato and
Lefever showed that $E_s$ is stable if $|E_s|<1$, but usually unstable
because of spontaneous pattern formation above that
threshold\cite{LL}.  We can now confess that Fig.~\ref{fig:jerry} was
generated by simulating the LL equation (for $\theta = 2.1$,
$E_{\mathrm in}(0)=1.5$ for {\sl (a)} and $2.5$ for {\sl (b)} and {\sl
(c)}), rather than the original full model of Moloney {\it et
al.}\cite{Mol}.  The strong similarity to the behaviour of the full
model shows how such mean-field models can capture the essential
features of a full cavity model while being both cheaper to simulate
and easier to analyse.

\begin{figure}[htbp]
\centering
    \includegraphics[width=0.75\textwidth]{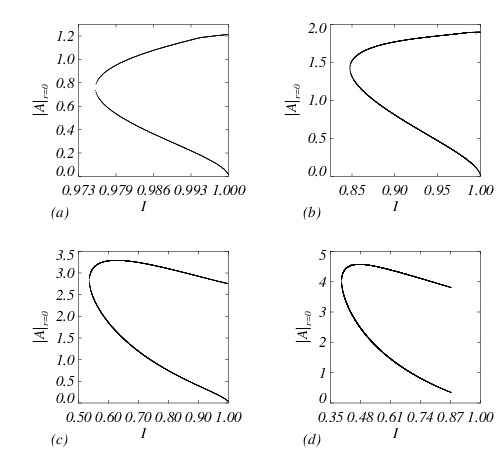}
\caption{ \label{fig:AovI}
Bifurcation diagrams illustrating the 2D Kerr cavity soliton states
for a) $\theta=0.9$, b) $\theta=1.2$, c) $\theta=1.5$ and d)
$\theta=1.8$.}
\end{figure}

Before developing the theme of mean-field models, it may be worthwhile
to describe an important result which shows that cavity solitons can
have properties beyond those of propagating spatial solitons.  It is
well known that the 2D NLS has a cylindrically symmetric soliton-like
solution $E(r,z)$, but that it is unstable, either diffracting away or
collapsing to a singularity\cite{Newell}.  What about the cavity
equivalent?  Firth and Lord\cite{KC} found stationary solutions to
the 2D LL in the form $E_s(1+A(r))$. For these cavity solitons $|A(0)|$
is plotted against $I=|E_s|^2$ in Fig.~\ref{fig:AovI}, for various
values of
$\theta$.  The solution curve is typical of cavity solitons,
forming a loop coexisting with the homogeneous background solution
(here $A=0$), in the range $I<1$ where the latter is stable.

\begin{figure}[htbp]
\centering
    \includegraphics[width=0.75\textwidth]{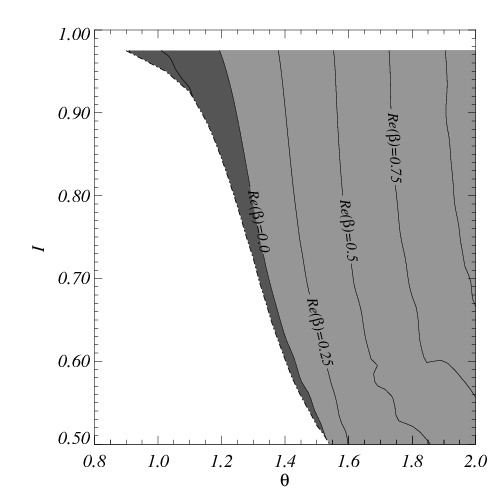}
\caption{ \label{fig:kerrstability} Existence domain of the 2D Kerr
cavity soliton (shaded region), stable only in the darker region (near
the existence boundary, c.f.  Fig.~\protect\ref{fig:AovI}).}
\end{figure}

Our present interest is in the stability of these cavity soliton
solutions.  As might be expected, the lower branch of the loop is
always unstable, but {\em the upper branch may be stable}\cite{PhSc}.
The domain of stability is shown in Fig.~\ref{fig:kerrstability}, where
$Re(\beta)$, the real part of the most unstable perturbation eigenvalue
of the upper-branch cavity soliton solution, is shown as a contour plot
in the ($\theta, I$) plane.  The stable domain $Re(\beta)<0$ is small,
vanishing as $\theta$ is increased - recall that the NLS limit is
$\theta \rightarrow\infty$.  Direct simulation confirms the stability
analysis.  A perturbed cavity soliton exhibits damped oscillations in
the stable domain, which become undamped as $\theta$ is increased and
the stability boundary crossed.  For large $\theta$ the cavity soliton
either decays or gets very narrow, suggestive of collapse\cite{KC},
much as the bulk material solitons do.

Thus we find stable 2D cavity solitons in the Lugiato-Lefever
mean-field model, a behaviour qualitatively different from its
bulk-medium equivalent.  Even if the stable domain is quite small, it
perhaps encourages us to look more closely at the LL equation, to see
which features we can vary, and which we should maintain, in exploring
the world of cavity solitons. Comparing our version of the LL equation
to the NLS, we note that the time derivative, and all three terms on
the right side, are features associated with the cavity, and thus ought
to be maintained.  This leaves the diffraction term, and the
nonlinearity, as terms we can play with.  We already modified the
diffraction term when we considered the 2D cylindrically symmetric Kerr
cavity solitons.  We can go further, by having a true 2D diffraction,
and can even add {\em dispersion} to make `3D' cavity
solitons\cite{TlidiPinos}.  Or we can replace diffraction with
dispersion, and consider e.g.  fibre cavities, where Mitschke and
co-workers\cite{Mit} have found evidence of soliton-like structures in
synchronously-pumped fibre loops, and Wabnitz\cite{Wab} has examined
data storage issues.  We have assumed a simple cavity with planar
mirrors: a nonlinear medium placed within a curved-mirror cavity can be
described using a generalised mean-field model involving terms in e.g.
$(x^2+y^2)E$, dependent on the ABCD matrix of the cavity\cite{ME}.  We
will not consider such cases\cite{Tar,Saff} here.

The nonlinearity offers great scope for variation.  Perhaps the obvious
generalisation from the Kerr nonlinearity is to a two-level
atom-like response, which becomes Kerr-like far from the atomic
resonance\cite{Roz,Tli,OBH}.  In particular, we will consider the
converse of the Kerr limit, namely exact atomic resonance, in which
the medium is simply a saturable absorber, with neither a nonlinear nor
even a linear refractive index contribution. Again we will find stable,
robust cavity solitons, again in a medium with no bulk spatial
solitons\cite{OBH}.  A further generalisation is to consider a nonlinearity
mediated by a material excitation. This opens up the further
possibility that the medium can have its own dynamics and spatial
(usually diffusive) coupling.  Semiconductors are particularly
interesting among such media, and it has been shown\cite{MilanPRL} that
cavity solitons extend to semiconductor models, even in the presence of
such `soliton-antagonistic' effects as diffusion and a measure of
self-defocusing.  Perhaps even more surprisingly, Michaelis {\it et al.}
\cite{JenaPRE} found bright solitons in a cavity model with a
purely-defocusing, diffusive saturable Kerr medium, such as is found in
semiconductors just below the band edge.  A different approach is to
couple several optical fields through, for example, a $\chi^{(2)}$
nonlinearity.  This has been shown in mean-field models to support
cavity solitons in both second-harmonic-generation (SHG)\cite{JenaPRL}
and optical parametric oscillator (OPO)\cite{Longhi,Stal,TMH}
configurations.

These cited examples may be just the beginning of an exploration of
cavity solitons in a wide range of configurations and media.  We will
not attempt to detail the cited papers, or to speculate on future
developments.  What may be worthwhile, however, is to discuss possible
physical interpretations of these structures, both because their
existence is sometimes counter-intuitive and as a guide to future
theory and experiments.  We have already considered `soliton-in-a-box'
and `self-trapped switching wave'. The other pictures mentioned in the
Introduction were  `pattern element', and `local nonlinear resonance'.
We now flesh out these pictures, using the saturable absorber cavity as
example system\cite{OBH}.

This system is modeled by a modified LL
equation in which the Kerr nonlinearity is replaced by a two-level
saturable absorption. This takes the form, after appropriate scaling:
\begin{eqnarray}
\frac{\partial E}{\partial t} &=&
E_{\mathrm in} -\left(1 +  i\theta +\frac{2C}{1 +\left| E \right|^2}\right) E  + i
\nabla_{\bot}^2 E.
\label{obh}
\end{eqnarray}
The new parameter $C$ can be regarded as the density of the saturable
medium, expressed in terms of absorptivity.  The field, and thus the
intensity $I$ of the homogeneous solution $E_s$, is scaled to the
saturation intensity. As is well known\cite{LALart} $E_s$ can become
unstable provided $C\ge4$, either through OB or, if transverse effects
are considered\cite{LOld}, through spatial modulational instability
leading to pattern formation\cite{FSepl}.  The transverse wave vector
$K_c$ of the favoured pattern obeys $K_c^2=-\theta$, which shows
immediately that patterns form only in a {\em detuned} cavity, and
furthermore only if the mistuning $\theta$ is {\em negative}.  It was
shown in\cite{FSepl} that this relation has the simple physical
interpretation that the pattern is a `tilted-wave resonance', in which
the transverse wave vector $K_c$ just compensates for the cavity
mistuning.  It was also shown that the favoured pattern is
hexagonal.

\begin{figure}[htbp]
\centering
    \includegraphics[width=0.75\textwidth]{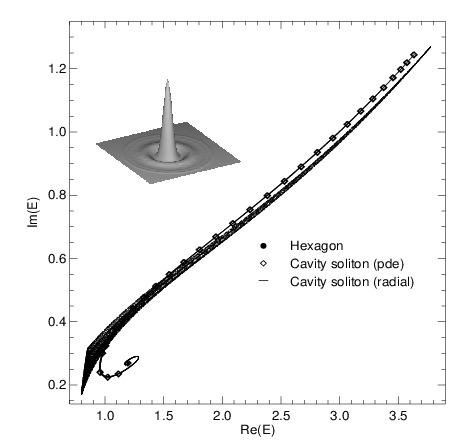}
\caption{ \label{fig:phaseplot}
Phase-plane plot of cavity soliton field (see text).}
\end{figure}

Now we consider cavity solitons in this model.  The profile of a cavity
soliton found by direct simulation of the pde~(4) is shown in
Fig.~\ref{fig:phaseplot}, inset to a plot of its field in the phase
plane (note that it lies wholly in the positive quadrant, so there is
no significant phase variation associated with the soliton). The
continuous line in Fig.~\ref{fig:phaseplot} is the same solution found
by direct radial integration of the time-independent equation for
$A(r)$.  These phase plots are seen to closely follow that of the
hexagonal pattern which coexists with the cavity soliton at these
parameters.  The plots deviate at the lower left, where the soliton
asymptotes to $E_s$ while the hexagon spike must match on to the
periodic hexagonal pattern.  Fig.~5 essentially defines the `pattern
element' interpretation: the soliton as an isolated hexagon spike.
Indeed in this example, and by experience in general, cavity solitons
are found only for parameter ranges where a stable pattern (usually
hexagonal in 2D) also exists. We should point out, however, that we
know of no analytic proof of this conjecture applicable to all the
cavity soliton models mentioned here.  Since switching waves can
involve patterns as well as homogeneous solutions, one could also
interpret this soliton as a self-trapped switching wave between the
hexagonal and the flat state.

There is a fourth interpretation, however, linked to the `tilted-wave'
interpretation of the patterns.  Suppose that the curvature of the
field profile at the soliton's centre allows the Laplacian term to
cancel the  mistuning term due to $\theta$.  Then the cavity is
(locally) brought into resonance, bleaching the medium, and making a
`hole' in the spatial profile of the absorption.  This is equivalent to
a gain guide, which confines the beam, activating the Laplacian term
$\nabla_\perp^2E$, and so on, forming a self-consistent feedback loop.
This  `local nonlinear resonance' interpretation seems to include some
aspects of the other three pictures: for example, one could view the
hexagonal pattern as a close-packed collection of local resonances.
Note that the local nonlinear resonance picture stresses key features:
the cavity (!), and the need for it to be off-resonance in the linear
limit.  It thus contrasts with `soliton-in-a-box', where the cavity
merely keeps the soliton supplied with medium. Our other two pictures
lie somewhere in between.

One way to investigate the relative validity and utility of these
pictures is to look at the perturbation response of cavity solitons.
We have already said that cavity solitons are stable, but the shape and
nature of their least stable eigenmodes should give information on their
nature {\em via} their response to perturbations.  For example, in a
`self-trapped switching wave' model of the soliton one might expect a
prominent eigenmode corresponding to un-trapping.

Investigations of these questions are in at a very early early stage.
Here we outline some preliminary results, again for the saturable
absorber cavity solitons, and only in one transverse dimension.  These
results may not be typical, of course, but we believe that they already
suggest that this approach has much to contribute to the physics and applications of cavity solitons.


\section*{Stationary Solutions and Stability}
For convenience, we rewrite equation (4) in terms of the variable $A$,
where $E=E_s(1+A)$:
\begin{eqnarray*}
\frac{\partial A}{\partial t} = i\nabla_\perp^2 A -(1+i\theta)A -
\frac{2C(1+A)}{1+I(1+A)(1+A^*)} +
  \frac{2C}{1+I}.
\end{eqnarray*}

We look for stationary solutions ($\partial_tA=0$) to this equation in
one spatial dimension, $\nabla_\perp^2 \rightarrow \partial_x^2$, by
discretising in space.  We thus obtain a large set of coupled nonlinear equations.  We use a Fast Fourier Transform (FFT) to compute the spatial derivatives in Fourier space.  This gives high accuracy (of order $N$, the 
number of spatial grid points).  From suitably chosen initial
conditions a Newton method is used to find solution(s) of the set.

The benefit of such a procedure over the more usual direct simulation
of the pde is that it can yield all the stationary solutions, not just
those which are dynamically stable.  Further, any solution's stability is
determined by the eigenvalues and eigenvectors of the Jacobian matrix
obtained by linearising around it, a matrix which is already calculated 
for use in the Newton method.  If any of the eigenvalues of the Jacobian 
have a positive real part then the solution is unstable.  The associated
eigenvector gives the spatial profile of the mode, whether stable,
unstable,or neutral.

\begin{figure}[htbp]
\centering
    \includegraphics[width=0.90\textwidth]{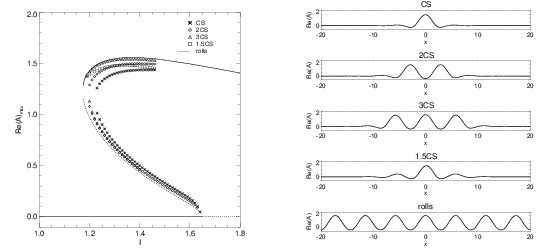}
\caption{ \label{fig:bifurcation}
A subset of the stationary solutions found for the parameters $C=5.4$,
$\theta=-1.2$, for which $E_s$ is single-valued.  The left hand panel
shows how maximum of the real part of the solution varies with
intensity $I$.  The right hand panel shows the spatial shape of the
solutions at $I=1.4$.}
\end{figure}

By this technique, we have computed a number of solutions and their
stability.  Some typical results are shown in
Fig.~\ref{fig:bifurcation}.
Of particular interest is the solution branch corresponding to a 
singly-peaked localised state (labelled $CS$), i.e. the cavity soliton. 
Just as for the Kerr cavity case in Fig.~\ref{fig:AovI}, this cavity soliton
co-exists with the homogeneous background solution $A=0$ and
the pattern state, in this case rolls (these would correspond to stripes
in 2D).  There also exist solutions
which are like multiple cavity solitons, $2CS$, $3CS$, etc.
\begin{figure}[htbp]
\centering
    \includegraphics[width=0.90\textwidth]{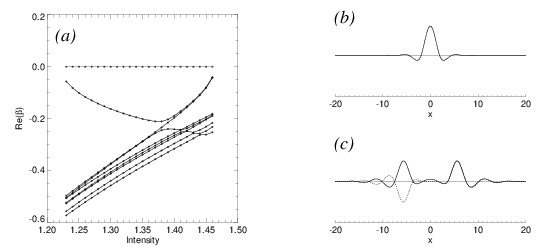}
\caption{ \label{fig:LSlambda}
{\sl (a)} For the stable segment of the {\sl LS} branch, the eight
eigenvalues with largest real parts as a function of the intensity
$I$.  {\sl (b)} At $I=1.23$, the eigenvector which becomes unstable for
smaller values of $I$.  {\sl (c)} At $I=1.46$, the pair of most
dangerous eigenvectors.}
\end{figure}

Turning now to the stability of the cavity soliton, as mentioned this
information is readily derived from the solution method used.  The
eigenfrequencies (imaginary parts of the eigenvalues) are responsible
for the response spectrum of the cavity soliton to perturbations.
Driving an eigenmode will make the cavity soliton oscillate at the
driving frequency, with an amplitude which will be maximum when the
driving field resonates with the mode, with the height and width of the
resonance determined by the real part of the eigenvalue.  For the
present problem, the spectrum is very simple: the modes with smallest 
damping all have {\em real} eigenvalues.  This means that, despite the
fact that the pump field $E_{\mathrm in}$ is detuned by $\theta$ from
the empty-cavity resonance, the perturbation response of the loaded
cavity is primarily at the pump frequency (nonlinear resonance).

We plot as a function of the intensity $I$ in Fig.~\ref{fig:LSlambda}
some of the  perturbation eigenvalues of the cavity soliton, to be
precise the eight  most `dangerous' ones, i.e. with least-negative
eigenvalues.  From the plot we see that the cavity soliton is stable
over the approximate range $1.22 < I < 1.46$.  Note that there always
exists a zero eigenvalue. It occurs because of the translational
symmetry in the $x$-direction.  Its existence is of great importance
for applications, as will be discussed in the next section.  As is
common for other solitons, this neutral mode has an eigenvector which
is the gradient of the state itself, because the gradient operator is
the generator of translations.

At the lower limit of this range in $I$, a second eigenvalue approaches
zero, and the cavity soliton loses stability at its existence limit
(see Fig.~\ref{fig:bifurcation}).  The corresponding eigenvector is
shown in panel {\sl (b)}.  It is highly localised on the cavity
soliton, and this mode thus strongly supports the picture of a cavity
soliton as a `local nonlinear resonance'.  As $I$ is increased, it can
be seen from Fig.~\ref{fig:LSlambda} that this mode becomes more damped
(but its form changes only slightly).  At about $I=1.38$ this
eigenvalue seems to undergo an anti-crossing with that of another
mode.  The latter mode, in conjunction with a third one, then assumes
dominance, until, at the upper limit of its range in $I$, the cavity
soliton loses stability to a degenerate pair of modes, with
eigenvectors shown as full and dotted lines in panel {\sl (c)}.  The
instability occurs where the {\sl CS} branch collides with the {\sl
1.5CS} branch.  These eigenvectors seem to be of switching-wave type,
being located where a neighbour soliton would be expected to appear.
This type of mode supports the `trapped switching wave' picture.
Incidentally,  the  modes associated with the other eigenvalues in
Fig.~\ref{fig:LSlambda} are extended, pattern-like states.  The damping
of such states is expected to weaken as the homogeneous solution
becomes less stable on approach to the MI threshold ($I=1.62$, see
Fig.~\ref{fig:bifurcation}) \cite{OBH}.

Thus the eigenmodes of the cavity soliton for this model seem to
suggest that it is primarily a local nonlinear resonance at low powers,
but that at the upper end of its range, it becomes more akin to the
self-trapped switching wave picture.  For applications, one might want
to work in the most stable region, where the two modes cross over.  At
that point, the correct interpretation is far from clear: trapped
switching wave? local nonlinear resonance? a bit of both?

It will be interesting to extend this approach to higher dimensions and
to other models.  In particular, cavity solitons in media with a
nonlinear index contribution can be expected to show a richer spectral
resonance behaviour, including in certain cases Hopf
bifurcations\cite{TMH,JenaVic}.


\section*{Applications}
Arrays of cavity solitons may have applications in parallel information
processing\cite{PASSrep}.  They can be created at any location by a
suitable address pulse, are non-diffracting and dynamically stable, all of
which makes them suitable `bits' for image or data capture, storage and
processing. This has been demonstrated numerically in \cite{OBH} and
confirmed in a prototype experiment\cite{PASSrep}.  A simple binary
memory array, such as implied by Fig.~\ref{fig:gun}, is unlikely to to
be competitive with electronic storage.  Cavity solitons, however, also
offer functionalities which are beyond any micro-structured material
array, whether optical or electronic.  In particular, they can be
optically manipulated, e.g. by imposing a spatial phase profile on the
driving field.  Processing schemes which take advantage of their unique
properties may avert the unequal competition with silicon which has
plagued other all-optical processing schemes.

The plasticity of cavity solitons can be linked to the `neutral mode'
identified in the previous section, which is associated with
the  translational invariance of the underlying equations
defining the cavity solitons.  Any perturbation to the pump field which
has a finite gradient at the soliton location will couple to the neutral

mode and cause the soliton to move.  The speed of the motion is
essentially proportional to this gradient, and so the soliton will
continue to move until it reaches a gradient-free location.  This has
implications for the response of the cavity solitons to noise and to
any stray gradients, and also for interaction between solitons.  Here
we will consider only its use to {\em control} the motion and location
of the solitons through the spatial phase profile of the pump field.
It is as though the solitons inhabit a `landscape' determined by the
phase of the holding (control) field, a landscape in which they move in
response to phase gradients. Thus a simple memory array\cite{OBH}
consists of a regular landscape of `hills' and `valleys', with the
solitons attracted to the peaks.  Unlike one formed from machined
pixels, however, this landscape is reconfigurable by changing the phase
profile of the control field. This allows cavity soliton bits to be
manipulated, by either global or local reconfigurations of the control
field.  No such manipulation is possible in material arrays, whether of
optoelectronic pixels or in silicon.  This plasticity opens up
possibilities for novel processing functions and applications such as a
`soliton carousel', a `soliton assembly-line processor' and a `zoom
memory'.  One can also expect useful applications in digital image
processing, for example in feature extraction.


\section*{Conclusion}
We have discussed a class of stable soliton-like structures predicted
to exist in driven optical cavities containing any of a wide variety of
nonlinear materials.  This class includes semiconductor
micro-resonators, which is promising for possible applications of these
cavity solitons.  They can be formed into two-dimensional arrays of
information bits which can be written, stored, read, erased \cite{EPL},
and spatially manipulated in various ways.  They can thus act as the
basis of a new kind of all-optical parallel processor, with
functionalities not available to other processing and storage devices
in information technology.

The first experimental verifications of these solitons have been made
\cite{PASSrep}, while they have  been found theoretically in quite a
wide variety of cavity systems containing a nonlinear optical medium.
We believe that in the coming years they will find an important role
both in optics and in optoelectronic technology.


\section*{Acknowledgements}
We thank Angus Lord and Andrew Scroggie for important contributions to
this paper, which is based on an invited talk delivered by WJF at the
Nonlinear Guided Waves '98 conference in Victoria, Canada, in April
1998.  WJF would like to thank the organisers and sponsors of that
meeting.  GKH acknowledges a grant from the RDF of the University of
Strathclyde.  We thank our partners in PASS, and its successor project
PIANOS, for many helpful discussions and insights.

\end{document}